\documentstyle[amssymb,preprint,aps]{revtex}
\begin{document}
\title{Energy-based theory of autoresonance phenomena: Application to
Duffing-like systems}
\author{Ricardo Chac\'{o}n}
\address{Departamento de Electr\'{o}nica e Ingenier\'{\i}a Electromec\'{a}%
nica,\\
Escuela de Ingenier\'{\i}as Industriales, Universidad de Extremadura,\\
Apartado Postal 382, E-06071\\
Badajoz, Spain}
\date{\today }
\maketitle

\begin{abstract}
A general energy-based theory of autoresonance (self-sustained resonance) in
low-dimensional nonautonomous systems is presented. The equations that
together govern the autoresonance solutions and excitations are derived with
the aid of a variational principle concerning the power functional. These
equations provide a feedback autoresonance-controlling mechanism. The theory
is applied to Duffing-like systems to obtain exact analytical expressions
for autoresonance excitations and solutions which explain {\it all} the
phenomenological and approximate results arising from a previous (adiabatic)
approach to autoresonance phenomena in such systems. The theory is also
applied to obtain new, general, and exact properties concerning
autoresonance phenomena in a broad class of dissipative and Hamiltonian
systems, including (as a {\it particular} case) Duffing-like systems.

\end{abstract}

It has been well known for about half a century that autoresonance (AR)
phenomena occur when a system continuously adjusts its amplitude so that its
instantaneous nonlinear period matches the driving period, the effect being
a growth of the system's energy. Autoresonant effects were first observed in
particle accelerators [1,2], and have since been noted in nonlinear waves
[3,4], fluid dynamics [5,6], atomic and molecular physics [7,8], plasmas
[9-11], nonlinear oscillators [12,13], and planetary dynamics [14-17].
Apparently, the first mention of the notion of resonance (\textquotedblleft
risonanza\textquotedblright ) was by Galileo [18]. Remarkably, this
linear-system-based concept has survived up to now: resonance (nonlinear
resonance) is identified with how well the driving period fits (a rational
fraction of) a natural period of the underlying conservative system [19].
However, the genuine effect of the frequency (Galilean) resonance (FR)
(i.e., the secular growth of the oscillation amplitude) can no longer be
observed in a periodically driven nonlinear system. As is well known, the
reason is simple: a linear oscillator has a single period which is
energy-independent, while nonlinear oscillators generally present an
infinity of energy-dependent periods. This means that, although an FR can
still be momentarily induced in a nonlinear system by exciting it with a
driving period that exactly matches the intrinsic period of the current
motion, the subsequent growth of the nonlinear oscillations changes the
intrinsic period of the motion, which no longer matches the excitation
period and thus takes the system out of FR. Since linear oscillations
represent a limiting {\it degenerate} (energy) case of the more general
nonlinear oscillations, it seems that any truly nonlinear generalization of
the notion of resonance, in its early (etymological) sense of {\it resonare}
(i.e., awaken an echo of some underlying nonlinear oscillation), should be
based on energy (or action) considerations. A case has been provided by the
notion of geometrical resonance [20]. Thus, if one is interested in
obtaining a nonlinear equivalent of the secular maintained growth intrinsic
to the FR, it is clear that the system must not be driven by a strictly
periodic excitation. In this regard, a previous theoretical approach to
autoresonance phenomena [3,7-11] provided an early explanation of the
mechanism inducing the growth of the oscillation (without the use of
feedback) for particular classes of resonantly driven nonlinear systems
which stay locked with an {\it adiabatically} varying perturbing oscillation
(the drive). The adiabatic excitation yields the autoresonant effect by
automatically adjusting the system's amplitude so that the instantaneous
nonlinear period matches the driving period. It should be stressed that a
fundamental part (hereafter referred to as adiabatic autoresonance (AAR)
theory, cf. refs. [9-11,21,22]) of the aforementioned previous theoretical
approach to AR phenomena presents severe limitations of applicability and
insight: essentially (see ref. [22] for a review), it was developed for
nonlinear oscillators that reduce to a Duffing oscillator 
\begin{equation}
\stackrel{..}{x}+\omega _{0}^{2}(x+bx^{3})=-\delta \stackrel{.}{x}%
+\varepsilon \cos \left( \omega _{0}t+\alpha t^{2}/2\right)  \eqnum{1}
\end{equation}%
for small amplitudes, where $\alpha $ is the {\it linear} sweep rate and $%
\delta >0$. In the context of AAR theory, it has been found {\it numerically}
that AR solutions only occur if (i) the damping coefficient $\delta $ is not
too{\it \ }large, and (ii) the amplitude of the AR oscillations grows on the
average, but also oscillates around the average growth. Also, AAR theory
predicts that (iii) there exists a threshold for AR, in particular, if the
normalized excitation amplitude $\varepsilon /\omega _{0}^{1/2}$ exceeds a
threshold proportional to $\alpha ^{3/4}$, the system will follow the
excitation to high amplitude, while the amplitude will stay very low
otherwise, (iv) that the threshold sweep rate $\alpha _{th}$ scales as $%
\delta ^{2}$, (v) that the AR effect is {\it solely} expected for the case
with initial conditions near some equilibrium of the (unperturbed) nonlinear
system, and (vi) that there exists a breaking time for AR, $t_{b}$.
Properties (ii), (iii), (v), (vi) also hold in (vii) the case with no
dissipation (cf. refs. [9, 10, 21]), but there has as yet been no
theoretical explanation of that fact. It is worth mentioning that, to the
best of the author's knowledge, the case of weak dissipation has only been
considered in a single previous work (cf. ref. [11]).

In this Letter a new, general, and energy-based theory for AR phenomena in
nonautonomous systems is presented and applied to the above Duffing
oscillators to explain conjointly points (i)-(vii) as well as to deduce new
properties concerning AR phenomena in generic systems (including
Duffing-like systems). The theory arises from the question as to whether
there exists an {\it upper} limit for the growth rate of the system's
amplitude when a {\it small}-amplitude force acts on the system. Consider
the general family of systems 
\begin{equation}
\stackrel{..}{x}\,=g(x)-d(x,\stackrel{.}{x})+p(x,\stackrel{.}{x})F(t), 
\eqnum{2}
\end{equation}%
where $g(x)\equiv -dV(x)/dx$ [$V(x)$ being an arbitrary time-independent
potential], $-d(x,\stackrel{.}{x})$ is a general damping force, and $p(x,%
\stackrel{.}{x})F(t)$ is an as yet undetermined suitable AR-inducing force.
Clearly, the corresponding equation for the energy is $\stackrel{.}{E}\,=\,%
\stackrel{.}{x}\left[ -d(x,\stackrel{.}{x})+p(x,\stackrel{.}{x})F(t)\right]
\equiv P(x,\stackrel{.}{x},t),$ where $E(t)\equiv (1/2)\stackrel{.}{x}%
^{2}(t)+V[x(t)]$ and $P(x,\stackrel{.}{x},t)$ are the energy and power,
respectively. In the spirit of the aforementioned energy-based approach to
resonance phenomena, the AR solutions are defined by imposing that the
energy variation $\Delta E=\int_{t_{1}}^{t_{2}}P(x,\stackrel{.}{x},t)dt$ is
a maximum (with $t_{1},t_{2}$ arbitrary but fixed instants), where the power
is considered as a functional. This implies a necessary condition (hereafter
referred to as the {\it AR condition}) to be fulfilled by AR solutions and
excitations, which is the Euler equation [23] 
\begin{equation}
\frac{\partial P}{\partial x}-\frac{d}{dt}\left( \frac{\partial P}{\partial 
\stackrel{.}{x}}\right) =0.  \eqnum{3}
\end{equation}%
From eq. (3), a relationship between $x,\stackrel{.}{x},$ and $F$ can be
deduced such that the solutions of the system given by eqs. (2) and (3)
together provide the AR excitations, $F_{AR}(t)$, and the AR solutions, $%
x_{AR}(t)$. It is worth noting that the AR condition (3) represents a {\it %
feedback} AR-controlling mechanism, which is absent in the aforementioned
previous approach to AR phenomena [3,7-11] where an explicit,
coordinate-independent, and adiabatic force is used from the beginning. In
this regard, autoresonant control has been previously discussed in the
context of vibro-impact systems [24] on the basis of the analysis of nearly
sinusoidal self-oscillations [25] where the term self-resonance was
introduced to indicate \textquotedblleft resonance under the action of a
force generated by the motion of the system itself\textquotedblright\ (cf.
ref. [25], p.166). The corresponding AR equations for the multidimensional
case can be straightforwardly obtained from the same principle and they will
be discussed elsewhere [26]. To compare the present approach with the
previous one [9,11,21,22] (cf. eq. (1)), consider the power functional $P(x,%
\stackrel{.}{x},t)=\stackrel{.}{x}\left[ -\delta \stackrel{.}{x}+F(t)\right] 
$. For the {\it particular} case of Duffing oscillators, the system (2), (3)
reduces to 
\begin{eqnarray}
\stackrel{..}{x}_{AR}+\omega _{0}^{2}\left( x_{AR}+bx_{AR}^{3}\right) 
&=&\delta \stackrel{.}{x}_{AR},  \eqnum{4a} \\
F_{AR} &=&2\delta \stackrel{.}{x}_{AR}.  \eqnum{4b}
\end{eqnarray}%
Note that eq. (4b) gives the AR condition, i.e., the AR excitations and the
(corresponding) AR solutions have the same instantaneous nonlinear period,
at all instants, but without the adiabaticity requirement of the AAR theory.
Generally, the AR condition (3) means that the instantaneous period of the
AR solution fits a rational fraction of the instantaneous period of the AR
excitation. This property generalizes (and contains as a particular case)
the persistent phase-locking condition of the AAR theory. To obtain AR
solutions (and hence AR excitations, cf. eq. (4b)) consider the ansatz{\it \ 
} $x_{AR}(t)=\gamma f(t)%
%TCIMACRO{\func{cn}}%
%BeginExpansion
\mathop{\rm cn}%
%EndExpansion
\left[ \beta g(t)+\phi ;m\right] $, where $%
%TCIMACRO{\func{cn}}%
%BeginExpansion
\mathop{\rm cn}%
%EndExpansion
$ is the Jacobian elliptic function of parameter $m$, and where the
constants $\beta ,m$, and the functions $f(t),g(t)$ have to be determined
for the ansatz to satisfy eq. (4a), while $\gamma ,\phi $ are arbitrary
constants. After substituting this ansatz into eq. (4a), one finds the exact
general AR solution 
\begin{eqnarray}
x_{AR}(t) &=&\gamma _{0}e^{\delta t/3}%
%TCIMACRO{\func{cn}}%
%BeginExpansion
\mathop{\rm cn}%
%EndExpansion
\left[ \varphi \left( t\right) ;1/2\right] ,  \nonumber \\
\varphi \left( t\right)  &\equiv &3\gamma _{0}\omega _{0}\sqrt{b}\left(
e^{\delta t/3}-1\right) /\delta +\varphi _{0},  \eqnum{5}
\end{eqnarray}%
with the constraint $\omega _{0}^{2}=2\delta ^{2}/9$ and where $\varphi
_{0}\equiv \phi +3\gamma _{0}\omega _{0}\sqrt{b}/\delta ,\gamma _{0}\equiv
\gamma $. Clearly, the exact AR excitation corresponding to solution (5) is 
\begin{equation}
F_{AR}(t)=\frac{2}{3}\gamma _{0}\delta ^{2}e^{\delta t/3}%
%TCIMACRO{\func{cn}}%
%BeginExpansion
\mathop{\rm cn}%
%EndExpansion
\left[ \varphi (t);1/2\right] -2\gamma _{0}^{2}\delta \omega _{0}\sqrt{b}%
e^{2\delta t/3}%
%TCIMACRO{\func{sn}}%
%BeginExpansion
\mathop{\rm sn}%
%EndExpansion
\left[ \varphi \left( t\right) ;1/2\right] 
%TCIMACRO{\func{dn}}%
%BeginExpansion
\mathop{\rm dn}%
%EndExpansion
\left[ \varphi \left( t\right) ;1/2\right] ,  \eqnum{6}
\end{equation}%
where $%
%TCIMACRO{\func{sn}}%
%BeginExpansion
\mathop{\rm sn}%
%EndExpansion
$ and $%
%TCIMACRO{\func{dn}}%
%BeginExpansion
\mathop{\rm dn}%
%EndExpansion
$ are the Jacobian elliptic functions. Observe that the particular
time-dependence of the AR solution (5) directly explains the above point
(ii) (see fig. 1). In comparing the present predictions with those from AAR
theory, recall that the latter {\it solely} exist for the case with $%
x(0)\simeq 0,$ $\stackrel{.}{x}(0)\simeq 0$, for $b>0$ (point (v)). Thus,
for this case $\gamma _{0}\simeq 0$ and hence eq. (6) can be approximated by 
$F_{AR}(t)\simeq \frac{2}{3}\gamma _{0}\delta ^{2}\left( 1+\frac{1}{3}\delta
t+...\right) 
%TCIMACRO{\func{cn}}%
%BeginExpansion
\mathop{\rm cn}%
%EndExpansion
\left[ \gamma _{0}\sqrt{b}\left( \omega _{0}t+\frac{1}{6}\omega _{0}\delta
t^{2}+...\right) ;1/2\right] $, and, using the Fourier expansion of $%
%TCIMACRO{\func{cn}}%
%BeginExpansion
\mathop{\rm cn}%
%EndExpansion
$ [27], one finally obtains 
\begin{equation}
F_{AR}(t)\simeq \frac{2}{3}\kappa \gamma _{0}\delta ^{2}\left( 1+\frac{1}{3}%
\delta t+...\right) \cos \left[ \kappa ^{\prime }\gamma _{0}\sqrt{b}\left(
\omega _{0}t+\frac{1}{6}\omega _{0}\delta t^{2}+...\right) \right] , 
\eqnum{7}
\end{equation}%
where $\kappa \equiv \pi \sqrt{2}%
%TCIMACRO{\func{csch}}%
%BeginExpansion
\mathop{\rm csch}%
%EndExpansion
(\pi /2)/K(1/2)\simeq 1,\kappa ^{\prime }\equiv \pi /(2K(1/2))\simeq 1$.
Now, one sees that to consider the excitation $\varepsilon \cos \left(
\omega _{0}t+\alpha t^{2}/2\right) $ (cf. eq. (1)) as a reliable
approximation to $F_{AR}(t)$ (cf. eq. (7)) implies that the damping
coefficient has to be sufficiently small (point (i)) so as to have a
sufficiently large breaking time, $t_{b}\sim \delta ^{-1}$ (point (vi)).
Thus, for $t\lesssim t_{b}$, one obtains $\varepsilon _{th}\sim \delta
^{2},\alpha _{th}\sim \omega _{0}\delta $ (cf. eqs. (1), (7)). When $\omega
_{0}\sim \delta $ (recall that $\omega _{0}^{2}=2\delta ^{2}/9$ for the
exact AR solution (5)), one finds $\alpha _{th}\sim \delta ^{2}$ (point
(iv)), {\it which explains the adiabaticity requirement of AAR theory for
dissipative systems}, $\varepsilon _{th}/\omega _{0}^{1/2}\sim \alpha
_{th}/\alpha _{th}^{1/4}\equiv \alpha _{th}^{3/4}$ (point (iii)), and the
cosine's argument in eq. (7) can be reliably approximated by the first two
terms, as in AAR theory (cf. eq. (1)). Figure 1 shows an illustrative
comparison between the AR responses yielded by AR excitations given by $%
\varepsilon \cos (\omega _{0}t+\alpha t^{2}/2)$, where in all cases $%
\varepsilon >\varepsilon _{th}$, and $F_{AR}(t)$ (cf. eq. (6)),
respectively, for the cases $\omega _{0}\sim \delta $ (fig. 1a) and $\omega
_{0}\gg \delta $ (fig. 1b). Point (vii) is rather striking in view of the
very different properties of Hamiltonian and dissipative systems, and its
explanation is a little more subtle. Firstly, note that current AR theory
provides an unsatisfactory result for the limiting Hamiltonian case. For
example, eq. (3) yields $r(x)\stackrel{.}{F}(t)=0$ for the family given by
eq. (2) with $d(x,\stackrel{.}{x})\equiv 0,p(x,\stackrel{.}{x})\equiv r(x)$,
i.e., including the cases of external and parametric (of a potential term)
excitations. Clearly, the two possible types of corresponding particular
solutions, equilibria and those yielded by a constant excitation (cf. eqs.
(2), (3)), can no longer be AR solutions. Secondly, for the above Duffing
oscillators one has $\stackrel{..}{x}_{AR}+\omega _{0}^{2}\left(
x_{AR}+bx_{AR}^{3}\right) =F_{AR}/2$ (cf. eq. (4)). Therefore, it is natural
to assume the ansatz $F(t)\equiv \lambda \stackrel{.}{x}(t),\lambda >0$, for
the case with no dissipation, where{\it \ }now the AR rate, $\lambda ,$ is a 
{\it free} parameter which controls the initial excitation strength. Thus,
the corresponding AR solutions are given by eq. (5) while AR excitations are
given by the expression in eq. (6) multiplied by 1/2, both with $\lambda $
instead of $\delta $, which explains point (vii) and hence {\it the
adiabaticity requirement of AAR theory for Hamiltonian systems} (recall
point (iv)). It is worth mentioning that this valuable result holds for the
broad family of dissipative systems $\stackrel{..}{x}+dV(x)/dx=-\delta 
\stackrel{.}{x}\left\vert \stackrel{.}{x}\right\vert ^{n-1}+F(t),$ where $%
V(x)$ is a generic time-independent potential and $-\delta \stackrel{.}{x}%
\left\vert \stackrel{.}{x}\right\vert ^{n-1}$ is a general dissipative force
($\delta >0,n=1,2,3,...$). The corresponding AR equations (cf. eqs. (2),
(3)) are $\stackrel{..}{x}_{AR}+dV(x_{AR})/dx_{AR}=n\delta \stackrel{.}{x}%
_{AR}\left\vert \stackrel{.}{x}_{AR}\right\vert ^{n-1}$, $F_{AR}=(n+1)\delta 
\stackrel{.}{x}_{AR}\left\vert \stackrel{.}{x}_{AR}\right\vert ^{n-1}$, and
hence one obtains $\stackrel{..}{x}_{AR}+dV(x_{AR})/dx_{AR}=nF_{AR}/(n+1).$
For the limiting Hamiltonian case ($\delta =0$), it is therefore natural to
assume the ansatz $F(t)\equiv n\lambda \stackrel{.}{x}\left\vert \stackrel{.}%
{x}\right\vert ^{n-1},\lambda >0$. Thus, AR solutions are the same for the
dissipative and Hamiltonian cases, while the AR excitations associated with
the Hamiltonian case are the (corresponding) AR excitations associated with
the dissipative case multiplied by $\frac{n}{n+1}$, with $\lambda $ instead
of $\delta $ for the Hamiltonian case [28]. In the light of the exact AR
excitation (cf. eq. (6)), one can readily obtain a reliable approximation
for {\it arbitrary} initial conditions, i.e., not just those near the
equilibrium of the unperturbed Duffing oscillator: 
\begin{eqnarray}
F_{AR}(t) &\simeq &\frac{2}{3}\kappa \gamma _{0}\delta ^{2}\left( 1+\frac{1}{%
3}\delta t+...\right) \cos \left[ \kappa ^{\prime }\gamma _{0}\sqrt{b}\left(
\omega _{0}t+\frac{1}{6}\omega _{0}\delta t^{2}+...\right) \right]  
\nonumber \\
&&-\kappa ^{\prime \prime }\gamma _{0}^{2}\delta \omega _{0}\sqrt{b}\left( 1+%
\frac{2}{3}\delta t+...\right) \sin \left[ \kappa ^{\prime }\gamma _{0}\sqrt{%
b}\left( \omega _{0}t+\frac{1}{6}\omega _{0}\delta t^{2}+...\right) \right] ,
\eqnum{8}
\end{eqnarray}%
where $\kappa ^{\prime \prime }\equiv \pi ^{2}\sqrt{2}%
%TCIMACRO{\func{sech}}%
%BeginExpansion
\mathop{\rm sech}%
%EndExpansion
\left( \pi /2\right) /K^{2}(1/2)\simeq 1.61819\simeq \left( 1+\sqrt{5}%
\right) /2$ (i.e., the golden ratio). Thus, for $t\lesssim t_{b}\sim \delta
^{-1}(\lambda ^{-1})$ one obtains the general (i.e., valid for any initial
condition) {\it 1st-order adiabatic} excitation 
\begin{equation}
F_{A,1}(t)=\varepsilon \cos \left( \omega _{0}t+\alpha t^{2}/2\right)
-\varepsilon ^{\prime }\sin \left( \omega _{0}t+\alpha t^{2}/2\right) , 
\eqnum{9}
\end{equation}%
with the above scalings for $\varepsilon _{th}$, $\alpha _{th}$, and $%
\varepsilon _{th}^{\prime }\sim \varepsilon _{th}\gamma _{0}b^{1/2}$. Figure
2 shows illustrative examples for several initial conditions far from $%
x\left( 0\right) =\stackrel{.}{x}\left( 0\right) =0$. Another fundamental
consequence of the present approach is the derivation of the scaling laws
for the thresholds corresponding to higher-order chirps [29]. Indeed,
consider the general n{\it th-order adiabatic excitation} $F_{A,n}(t)\equiv
\varepsilon \cos \left[ \omega \left( t\right) t\right] -\varepsilon
^{\prime }\sin \left[ \omega \left( t\right) t\right] $, $\omega \left(
t\right) \equiv \omega _{0}+\sum_{n=1}^{\infty }\alpha _{n}t^{n},$ instead
of $\varepsilon \cos \left( \omega _{0}t+\alpha t^{2}/2\right) $ in eq. (1),
where $\alpha _{n}$ is the nth-order sweep rate $\left( \alpha _{1}\equiv
\alpha /2\right) $. For this general case, the above analysis
straightforwardly yields the scaling law $\varepsilon _{th}/\omega
_{0}^{1/2}\sim N(n)\alpha _{n,th}^{3/(2n+2)}$ for $t\lesssim t_{b}\sim
\delta ^{-1}(\lambda ^{-1})$, where $\alpha _{n,th}$ is the threshold
nth-order sweep rate and $N(n)\equiv \left[ 3^{n}\left( n+1\right) !\right]
^{3/\left( 2n+2\right) }$ is a monotonous increasing function. Thus, the $3/4
$ scaling law is a particular law which solely applies to a linear chirp.
For the case of a single chirp term $\left( \omega \left( t\right) \equiv
\omega _{0}+\alpha _{n}t^{n}\right) $, the dependence of the above general
scaling law on $n$ indicates that one can expect a similar AR effect for 
{\it ever smaller} values of $\alpha _{n}$ as $n$ increases. Computer
simulations confirm this point: an illustrative example is shown in fig. 3.

A further question remains to be discussed: We have seen why AAR theory
requires AR excitations to be adiabatically varying perturbing oscillations,
but which are the underlying adiabatic invariants? To answer this question,
note that eq. (4a) (with $\lambda $ instead of $\delta $ for the case with
no dissipation) can be derived from a Lagrangian, which one defines as $%
L=e^{-\delta t}\left( p^{2}/2-\omega _{0}^{2}x^{2}/2-\omega
_{0}^{2}bx^{4}/4\right) ,$ $p\equiv \stackrel{.}{x}$, and whose associated
Hamiltonian is $H=p^{2}e^{\delta t}/2+\omega
_{0}^{2}(x^{2}/2+bx^{4}/4)e^{-\delta t}.$ The form of this Hamiltonian
suggests the following simplifying canonical transformation: $X=xe^{-\delta
t/2},P=pe^{\delta t/2}.$ It is straightforward to see that the generating
function of the canonical transformation [30] is $F_{2}(x,P,t)=xPe^{-\delta
t/2}$. The new Hamiltonian therefore reads: $K(X,P,t)=H(x,p,t)-\partial
F_{2}/\partial t=P^{2}/2+\omega _{0}^{2}(X^{2}/2+be^{\delta
t}X^{4}/4)+\delta PX/2.$ In the limiting linear case $\left( b=0\right) $,
one sees that $K$ is conserved, i.e., the AR solutions corresponding to the
linear system are associated (in terms of the old canonical variables) with
the invariant{\it \ } $e^{\delta t}p^{2}/2+\omega _{0}^{2}e^{-\delta
t}x^{2}/2+\delta xp/2$, while for the nonlinear case ($b\neq 0$) one obtains
(after expanding $e^{\delta t}$) that the respective AR solutions are
associated with the adiabatic invariant{\it \ } $p^{2}/2+\omega
_{0}^{2}\left( x^{2}/2+bx^{4}/4\right) +\delta xp/2\equiv E+\delta xp/2$
over the time interval $0\leqslant t\leqslant t_{AI}$, $t_{AI}\sim \delta
^{-1}$ (i.e., the same scaling as for the breaking time, $t_{b}$, deduced
above), where $E$ is the energy of the underlying integrable Duffing
oscillator. Observe that the adiabatic invariant reduces to $E$ provided
that $\delta $ $\left( \lambda \right) $ is sufficiently small (as required
in AAR theory) and that the {\it same} result is obtained for a general
potential $V(x)$ instead of Duffing's potential.

In sum, a general energy-based theory of AR phenomena in low-dimensional
nonautonomous systems has been deduced from a simple variational principle
concerning the power functional. In particular, the theory explains all the
phenomenological and approximate results arising from a previous adiabatic
approach to AR in Duffing-like systems. For this class of systems, the
present theory also explains the adiabaticity requirement as well as why the
same theoretical predictions hold in the cases with and without dissipation,
and yields the analytical expression for the adiabatic invariants.
Additionally, new adiabatic approximations to AR excitations are derived
concerning two general cases which were not considered in the previous
adiabatic approach, namely, the case of arbitrary initial conditions (not
just those near equilibria) and the case of arbitrary potential (not just
linear) chirps, for which new general scaling laws were deduced (including
the 3/4 scaling law as a particular case). Computer simulations confirmed
all the theoretical predictions. In view of the generality of the present
theory of AR, one can expect it to be quite readily testable by experiment
(e.g., in the Diocotron system in pure-electron plasmas), and that it will
find applications in different fields of physics, such as plasmas, fluids,
and solar system dynamics.

The author thanks Professor Friedland for kindly providing a reprint of ref.
[22], which was the origin of the present work, and Professors Gallas,
Goldhirsch, Malhotra, Malomed, and S\'{a}nchez for useful comments. The
author acknowledges financial support from Spanish MCyT through
FIS2004-02475 project.

\bigskip

\bigskip

\bigskip

\bigskip

\bigskip

\bigskip

\bigskip

\bigskip

\bigskip

\bigskip

\bigskip

\bigskip

\subsection{Figure Captions}

Figure 1. Autoresonant responses (energy vs time, both variables in
arbitrary units) to a linearly swept excitation (cf. eq. (1), grey lines)
and to an exact AR excitation (cf. eq. (6), black lines), for the parameters 
$b=5,x(0)=10^{-3},\stackrel{.}{x}(0)=0,\gamma _{0}=10^{-3},\varphi _{0}=0$.
(a) Case with $\omega _{0}\sim \delta ,$ as required for an exact AR
excitation, and $\varepsilon =0.5$. (b) Case with $\omega _{0}\gg \delta $,
i.e., far from the exact AR excitation requirement, and $\omega _{0}=2\pi $.

Figure 2. Autoresonant responses (energy vs time, both variables in
arbitrary units) to a general 1st-order adiabatic excitation (cf. eq. (9),
black lines) and to a harmonic and linearly swept excitation (cf. eq. (1),
grey lines), for the parameters $b=5,\delta =0.4,\omega _{0}=0.2,\varepsilon
=0.5\sim \varepsilon _{th},\alpha =0.08\sim \alpha _{th}$, and initial
conditions $x(0)=0.8,\stackrel{.}{x}(0)=0.107$ ($\gamma _{0}=0.8,\varepsilon
^{\prime }=0.9\sim \varepsilon ^{\prime }{}_{th}$, thick lines) and $%
x(0)=0.6,\stackrel{.}{x}(0)=0.08$ ($\gamma _{0}=0.6,\varepsilon ^{\prime
}=0.67\sim \varepsilon _{th}^{\prime },$ thin lines).

Figure 3. Autoresonant responses (energy vs time, both variables in
arbitrary units) to a harmonic excitation with a linear chirp ($\omega
(t)=\omega _{0}+\alpha _{1}t$, cf. eq. (1), grey lines) and with a quadratic
chirp ($\omega \left( t\right) =\omega _{0}+\alpha _{2}t^{2}$, black lines),
for the parameters $b=5,\delta =0.4,\omega _{0}=0.2,\varepsilon =0.5\sim
\varepsilon _{th},\alpha _{1}=0.04\sim \alpha _{1,th},\alpha _{2}=0.003\sim
\alpha _{2,th}$, and the initial conditions $x(0)=10^{-3},\stackrel{.}{x}%
(0)=0$ (thick lines) and $x(0)=0,\stackrel{.}{\stackrel{.}{x}(0)=1\text{
(thin lines).}}$

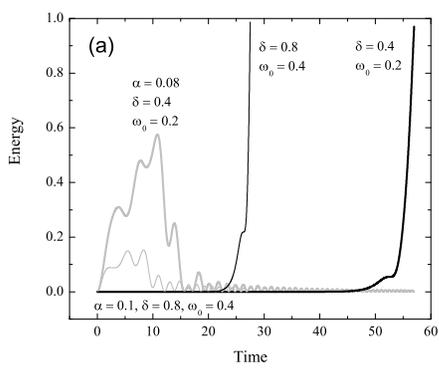
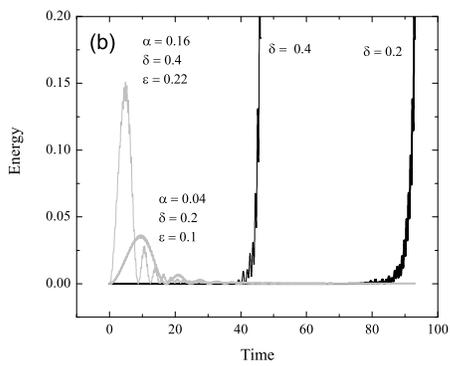
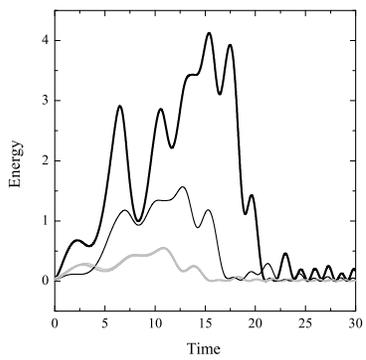
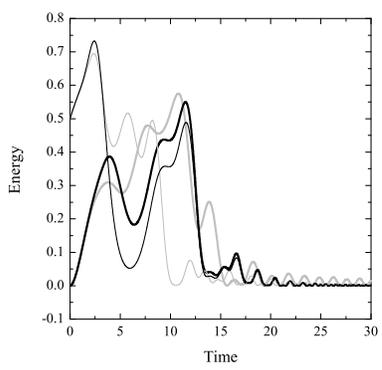
\end{document}